\documentclass[10pt]{iopart}
\usepackage{amsfonts,amssymb,mathrsfs}

\newcommand{\be}{\begin{equation}}
\newcommand{\ee}{\end{equation}}
\newcommand{\bea}{\begin{eqnarray}}
\newcommand{\eea}{\end{eqnarray}}

\begin{document}
\title{Clifton's spherical solution in $f(R)$ vacuo harbours a 
naked singularity}
\author{Valerio Faraoni}
\address{Physics Department, Bishop's University, 
2600 College St., Sherbrooke, 
Qu\'{e}bec, Canada J1M~1Z7}
\eads{\mailto{vfaraoni@ubishops.ca}}
\date{\today} 

\begin{abstract} 
Clifton's exact solution of $f(R)=R^{1+\delta}$ gravity 
describing a dynamical spherical metric which is asymptotically 
Friedmann-Lemaitre-Robertson-Walker is studied. It is shown that 
it harbours a strong spacetime singularity at a finite radius and 
that this singularity is naked.
\end{abstract} 

\section{Introduction}

Type Ia supernovae have provided us with  the knowledge that  
the universe is currently in 
a phase of 
accelerated expansion \cite{SN}. This acceleration  has been 
modelled in various ways; the most common models are probably 
dark energy ones based on General 
Relativity (hereafter GR, see 
\cite{Linderresletter} for a list of references). However, the 
exotic and {\em ad hoc} dark energy leaves many cosmologists 
dissatisfied and attempts have been made to model the cosmic 
acceleration without dark energy. $f(R)$ 
theories of gravity akin to the quadratic theories required by 
the renormalization of GR  have been introduced in the metric 
\cite{CCT}, Palatini \cite{Vollick}, 
and metric-affine \cite{metricaffine} formulations and have 
received  much attention in recent years (see \cite{review} for a 
review  and \cite{otherreviews} for short introductions). 

Along with cosmological and other considerations ({\em e.g.}, 
stability, weak-field limit, ghost content), it is important to 
understand spherically symmetric solutions in these theories, 
a task which has proved to be non-trivial (see \cite{Frolovetc} 
and 
references therein). For definiteness,  we consider here metric 
$f(R)$ gravity described by the action
\be
S=\frac{1}{2\kappa} \int d^4x \sqrt{-g}\, f(R) +S^{matter} \;,
\ee
where $f(R)$ is a non-linear function of its argument and  
$S^{matter} $ is the matter part of the action. $R$ 
denotes  the Ricci scalar of the metric $g_{ab}$ with determinant 
$g$, $\kappa=8\pi G$ where $G$ is Newton's constant, and we 
adopt  the notations of Ref.~\cite{Wald}. 

It is well known that the Jebsen-Birkhoff 
theorem does not hold in these theories, which adds to the 
richness and variety of spherically symmetric solutions. Of 
particular interest are black holes in generalized gravity, which 
have been studied especially in relation to their  
thermodynamics~\footnote{The thermodynamics of local Rindler 
horizons in $f(R)$ gravity, which is used to derive the 
classical field equations as  an equation of state 
\cite{Eling} is modeled after the thermodynamics of dynamical 
$f(R)$ black 
holes.} ({\em e.g.}, \cite{Brusteinetal}).  Since $f(R)$ theories 
are designed to produce a time-varying effective cosmological 
constant, the spherically symmetric and black hole solutions of 
interest likely represent central objects embedded in 
cosmological 
backgrounds. Not much is known about this kind of objects even in 
the context of GR, although a few examples are available 
\cite{SultanaDyer, McClureDyer, FaraoniJacques,  
SaidaHaradaMaeda, Gaoetal, Hideki, 
HidekiHaradaCarr, fate} in Einstein's theory. Even less is known 
about $f(R)$ black holes and spherically symmetric solutions, 
which 
deserve to be understood better. Here we consider a specific 
solution proposed in $f(R)=R^{1+\delta}$ 
gravity in \cite{Clifton}.
The observational constraints set the  limits $\delta 
=\left( -1.1\pm 1.2 \right) \cdot 10^{-5}$ on the parameter $ 
\delta$ \cite{Clifton, BarrowClifton}, while local 
stability requires $f''(R) 
\geq 0$ \cite{DolgovKawasaki, mattmodgrav}, {\em i.e.}, $\delta 
>0$ hence we will use positive values of this parameter. 

The 
solution proposed in \cite{Clifton} is dynamical and presumably 
represents some kind of  dynamical central object embedded in a 
spatially flat FLRW  background in vacuum $f(R)=R^{1+\delta}$ 
gravity. This solution is made possible by the fact that  the 
fourth order field equations of vacuum metric $f(R)$ gravity
\be
f'(R)R_{ab}-\frac{f(R)}{2}\, g_{ab}=\nabla_a\nabla_b 
f'(R)-g_{ab} \Box f'(R) 
\ee
can be rewritten in the form of effective Einstein equations with 
geometric terms acting as a form of effective matter as
\be 
R_{ab}-\frac{1}{2}\, g_{ab} R=\frac{1}{f'(R)}\left[ 
\nabla_a\nabla_b f'-g_{ab} \Box f' +g_{ab}\frac{\left( 
f-Rf'\right)}{2} \right] \;.
\ee
In this picture the effective matter spoils the Jebsen-Birkhoff 
theorem and fuels the cosmic acceleration. Alternatively, an 
equivalent representation of $f(R)$ gravity as a Brans-Dicke 
theory with a scalar field potential exhibits a massive spin zero   
degree of freedom that causes these effects \cite{review}. 
Since exact spherically symmetric dynamical solutions of $f(R)$ 
gravity in asymptotically FLRW backgrounds are harder to find 
than in  GR (where few are known anyway) and are 
therefore valuable, we study Clifton's solution in the 
following.

\section{Clifton's  spherically symmetric dynamical solution}

Clifton's spherically symmetric dynamical solution in vacuum 
$f(R)=R^{1+\delta}$ gravity \cite{Clifton} is given by
\be\label{1}
ds^2=-A_2(r)dt^2+a^2(t)B_2(r)\left( dr^2 +r^2 d\Omega^2 \right) 
\;,
\ee
where $d\Omega^2= d\theta^2 +\sin^2 \theta \, d\varphi^2$ is the 
line element on the unit 2-sphere,
\begin{eqnarray}
A_2(r) &=& \left( \frac{1-C_2/r}{1+C_2/r}\right)^{2/q} \;, 
\label{2} \\
&&\nonumber \\
B_2(r) &=& \left( 1+\frac{C_2}{r} \right)^{4}A_2(r)^{\, q+2\delta 
-1} 
\;,\label{4}\\
&&\nonumber \\
a(t) &= & t^{\frac{ \delta ( 1+2\delta)}{1-\delta}} \;,\label{3}\\
&&\nonumber \\
q^2 &= & 1-2\delta+4\delta^2 \;, \label{5}
\end{eqnarray}
in isotropic coordinates and using the notation of 
\cite{Clifton} for the metric components. 
The line element~(\ref{1}) reduces to the 
Friedmann-Lemaitre-Robertson-Walker one in the limit 
 $C_2\rightarrow 0$. In the 
limit $\delta \rightarrow 0$ in which the theory reduces to 
GR, the metric~(\ref{1}) reduces to the 
Schwarzschild solution in isotropic coordinates. This suggests 
that  the positive root be taken in the expression $q=\pm 
\sqrt{1-2\delta +4\delta^2}$  deriving from eq.~(\ref{5}), and 
that $ q\simeq 1-\delta$ in the limit $|\delta |<<1$. Moreover, 
only positive values of the constant $C_2$ will be considered 
since the latter reduces to the Schwarzschild mass in the limit 
to GR. The solution~(\ref{1})-(\ref{5}) is conformal to the 
Fonarev 
solution \cite{Fonarev}  which is conformally static 
\cite{HidekiFonarev}, and therefore is also conformally static. 
This 
is a property shared with the Sultana-Dyer solution 
\cite{SultanaDyer} and with some representatives of 
the class of generalized McVittie 
solutions \cite{fate}.

We now want to write the metric~(\ref{1}) in the Nolan gauge, in 
which it is straightforward to identify the apparent horizons 
(if they exist). To this end, we make use of the 
Schwarzschild-like radial coordinate
\be \label{6}
\tilde{r} \equiv r \left( 1+\frac{C_2}{r} \right)^2 \;,
\ee
in terms of which $dr=\left( 1-\frac{C_2^2}{r^2} \right)^{-1} 
d\tilde{r} $ and we eventually transform to the areal radius
\be\label{8}
R \equiv \frac{ a(t) \sqrt{B_2(r)} \, \tilde{r} }{\left( 
1+\frac{C_2}{r} \right)^2} =a(t) \, \tilde{r} \, A_2(r)^{ 
\frac{q+2\delta -1}{2}}  \;.
\ee
The line element~(\ref{1}) then becomes
\be\label{9}
ds^2=-A_2dt^2 +a^2 A_2^{2\delta -1}d\tilde{r}^2 +R^2 d\Omega^2 
\;.
\ee
Using the fact that
\be \label{10}
d\tilde{r}=\frac{
dR-A_2^{\frac{q+2\delta -1}{q} } \dot{a} \, \tilde{r} \, dt}{
a\left[ A_2^{\frac{q+2\delta -1}{2}} +\frac{2( q+2\delta -1)}{q} 
\frac{C_2}{\tilde{r}} A_2^{\frac{2\delta -1-q}{2}} \right]} 
\equiv 
\frac{ dR-A_2^{\frac{q+2\delta-1}{q}} \dot{a} \, \tilde{r} \, dt 
}{a 
A_2^{\frac{q+2\delta-1}{2}} C(r)} \;,
\ee
where an overdot denotes differentiation with respect to $t$ and 
\be\label{11}
C(r)=1+\frac{2(q+2\delta -1)}{q} \, \frac{C_2}{\tilde{r}} \, 
A_2^{-q} = 
1+\frac{2(q+2\delta -1)}{q} \, \frac{C_2 a}{R} \, 
A_2^{\frac{2\delta -1-q}{2}} 
\ee
the metric assumes the Painlev\'e-Gullstrand-like form
\begin{eqnarray}
ds^2 & = & - A_2\left[ 1-\frac{A_2^{\frac{q+2(2\delta -1)}{q}}}{
A^qC^2}\, \dot{a}^2 \tilde{r}^2 \right] 
dt^2-\frac{2A^{\frac{q+2\delta-1}{q}}}{A^qC^2} \, 
\dot{a} \, \tilde{r} \, dtdR \nonumber\\
&&\nonumber\\
&+ &  \frac{dR^2}{A^q C^2}+R^2d\Omega^2 \;.\label{12}
\end{eqnarray}
In order to eliminate the cross-term in $dtdR$ we introduce the 
new time coordinate $\bar{t}$ defined by
\be\label{13}
d\bar{t}=\frac{1}{F(t,R)} \left[ dt +  \beta(t,R)dR \right] \;,
\ee
where $F(t,R)$ is an integrating factor that satisfies the 
equation
\be\label{14}
\frac{\partial}{\partial R}\left( \frac{1}{F} \right)=
\frac{\partial}{\partial t}\left( \frac{\beta}{F} \right)
\ee
to ensure that $d\bar{t}$ is an exact differential.  The 
line element then becomes 
\begin{eqnarray}
ds^2 & = & -A_2\left[ 1-\frac{A_2^{\frac{q+2(2\delta -1)}{q}}}{ 
A^qC^2}\,  \dot{a}^2 \tilde{r}^2 \right] F^2 d\bar{t}^2 
\nonumber\\
&&\nonumber\\
 &+ & 2F  
\left\{ A_2\beta \left[  1-\frac{A_2^{\frac{q+2(2\delta -1)}{q 
}}}{A^qC^2}\, \dot{a}^2 \tilde{r}^2 \right]
-1-\frac{A_2^{\frac{q+2\delta-1}{q}} }{A^qC^2} \, 
\dot{a}  \tilde{r} \right\}
d\bar{t}dR \nonumber\\
&&\nonumber\\
& + & \left\{-A_2 
\left[ 1-\frac{A_2^{\frac{q+2(2\delta -1)}{q}}}{A^qC^2}\, 
\dot{a}^2  \tilde{r}^2 \right]\beta^2 
+\frac{2A_2^{\frac{q+2\delta -1}{q}}}{A^q C^2}\, 
\dot{a}\tilde{r}\beta +\frac{1}{A^qC^2} \right\}\nonumber\\
&&\nonumber\\ 
& + &R^2d\Omega^2 
\;.\label{15}
\end{eqnarray}
By setting 
\be\label{16}
\beta= \frac{ A_2^{ \frac{q+2\delta-1}{q} } }{A_2^q C^2}\, 
  \frac{\dot{a} \, \tilde{r}}{ A_2\left[ 1-
\frac{A_2^{\frac{q+2(2\delta -1)}{q}} }{A_2^q C^2}\, 
\dot{a}^2 \tilde{r}^2 \right]}  
\ee
the $dtdR$ cross-term disappears and we are left with the Nolan 
gauge metric 
\begin{eqnarray}
ds^2 & = & -A_2 D F^2 d\bar{t}^2 +\frac{1}{A_2^q C^2} \left[
1+\frac{ A_2^{ \frac{ 2q(1-q)+2\delta(2-q)-2}{q} } H^2R^2}{C^2 D} 
\right] dR^2 \nonumber\\
&&\nonumber\\
&+ & R^2 d\Omega^2 \;, \label{18}
\end{eqnarray}
where $ H\equiv \dot{a}/a$ and 
\be\label{17}
D\equiv 1-\frac{ A_2^{\frac{q+2(2\delta -1)}{q} } }{A_2^q C^2}\, 
\dot{a}^2\tilde{r}^2 =
1-  \frac{ A_2^{\frac{q(2-q) +2\delta(2-q) -2}{q} } }{A_2^q 
C^2}\, 
H^2 R^2 \;.
\ee
The apparent horizons, if they exist, are located at $g^{RR}=0$. 
This equation is satisfied if  $A_2=0$, or $ D=0$, or $C=0$,   
corresponding to 
\begin{eqnarray}
&& r=  C_2 \;, \label{19}\\
&&\nonumber \\
&& A_2^q C^2 =A_2^{\frac{q(2-q)+2\delta (2-q)-2}{q}} H^2R^2 \;, 
\label{20}\\
&&\nonumber \\
&& 1+\frac{ 2(q+2\delta -1)}{q}\, \frac{C_2A}{R}\, 
A_2^{\frac{2\delta -1-q}{2}}=0 \;, \label{21}
\end{eqnarray}
respectively. The locus $r=C_2$ for which 
$A_2=0$ (which describes the 
Schwarzschild  horizon in the limit $\delta\rightarrow 
0$ in which the theory reduces to GR)  corresponds 
to a spacetime singularity. In fact, the Ricci scalar is
\be\label{22}
{R^a}_a=\frac{ 6\left( \dot{H}+2H^2 \right)}{A_2(r)} 
\ee
and diverges as $r\rightarrow C_2$ (it reduces 
to the familiar value $6\left( \dot{H}+2H^2 \right)$ in the 
$C_2\rightarrow 0$ limit). Furthermore, this singularity is a 
strong one in the sense of Tipler's classification \cite{Tipler}: 
the metric determinant is
\be\label{23}
g=-a^6(t)r^4 \left( 1+\frac{C_2}{r}\right)^{12} 
A_2(r)^{3q+6\delta-2} 
\ee
and vanishes as $r\rightarrow C_2$. The volume of a body is 
shrunk  to zero as it approaches this singularity and the 
energy density  of a (real or effective) fluid diverges there. No 
object can  cross the locus $r=C_2$. The regions $0< r< C_2$ and 
$r>C_2$ describe two disconnected 
spacetimes. It seems that the pull of the effective matter in the 
universe has stretched the $r=0$ singularity of the Schwarzschild 
black hole into a sphere.~\footnote{This feature is consistent 
with the known phenomenology of the Sultana-Dyer solution, of  
generalized  McVittie  solutions \cite{VFinpreparation}, and of 
higher-dimensional Gauss-Bonnet black holes \cite{Hideki}.}

Eq.~(\ref{21}), corresponding to $C=0$, has no solutions for 
$\delta>0, C_2>0$, and $ R>0$. In fact, for $ 0< \delta <<1$, 
this 
reduces to $ 1+2\delta \, \frac{C_2}{r\left( 1-\frac{C_2}{r} 
\right)^2}=0$, which cannot be satisfied if $ C_2 r \delta >0$.   
Let us focus on eq.~(\ref{20}) corresponding to 
$D=0$. This yields
\be\label{24}
HR=\pm\left[ 1+\frac{2(q+2\delta -1)}{q}\, \frac{C_2 a}{R}\, 
A_2^{\frac{2\delta-1-q}{2}} \right] A_2^{\frac{ 
2q^2-2q-4\delta+2\delta q+2}{q}} \;.
\ee
In  the limit of small $\delta$ this equation reduces to $HR=\pm 
\left[ 
1+\frac{2\delta C_2 a}{R}\, A_2^{-\left(1-\frac{3\delta}{2} 
\right)} \right] A_2^{1-\delta} $ and, in an expanding universe 
in which $HR \geq 0$, 
we discard the negative sign in eq.~(\ref{24}). The apparent 
horizons, if they exist, are located at the roots of the equation
\begin{eqnarray}
&& HR^2 -A_2^{\frac{2q^2-2q-4\delta+2\delta q +2}{2q}} \, R 
-\frac{  2(q+2\delta-1)}{q} \, C_2 a A_2^{\frac{ 
2q^2-2q-4\delta+2\delta 
q+2}{2q}+\frac{2\delta -1-q}{2} } \nonumber\\
&&\nonumber\\
&& =0 \;.
\label{25}
\end{eqnarray}
Although eq.~(\ref{25}) is written in the form of a 
quadratic algebraic 
equation, it is really an implicit equation for the 
$R$-coordinate of the apparent horizons because the coefficients 
are functions of $r(R)$. In spite of this fact, it is still 
useful to regard eq.~(\ref{25}) as a formal algebraic equation.

To gain some insight, consider the following two limits. In the 
limit $C_2\rightarrow  0$ in which the central object disappears 
and the solution is a FLRW space, $r=\tilde{r}$ becomes a 
comoving radius  and $R$ becomes a proper radius, while 
eq.~(\ref{25}) reduces to 
\be
R\left( HR-1\right)=0
\ee
which yields as a solution $R_{c}=1/H$, the radius of the
cosmological horizon.

In the limit $\delta\rightarrow 0 $ in 
which the theory reduces to GR, the exponent
\be\label{26}
\frac{2q^2-2q-4\delta +2\delta q +2}{2q} 
\approx 1-\delta   
\rightarrow 1 \;,
\ee
while $\frac{-2(q +2\delta -1)}{q} \approx -2\delta \rightarrow 
0$, $H\rightarrow 0$ and eq.~(\ref{25}) simply yields $ A_2 R=0$, 
or 
$ a\tilde{r}A_2^{\frac{q+2\delta +1}{2}}=0 $ and $A_2=0$, 
hence $r=C_2$.  This  is the usual Schwarzschild horizon 
expressed using the isotropic radius (the corresponding 
Schwarzschild radius is $\tilde{r}=2C_2$).

Returning to the general case, we see that eq.~(\ref{25}) has the 
formal solutions
\begin{eqnarray} 
 R_{1,2} & = & \frac{A_2^{\frac{2q^2-2q-4\delta+2\delta q+2}{2q}} 
\pm 
\sqrt{\Delta}}{2H} \nonumber\\
&&\nonumber\\
&=& 
\frac{A_2^{\frac{2q^2-2q-4\delta+2\delta q+2}{2q}} }{2H}
\left[ 1\pm \sqrt{ 1+\frac{8(q+2\delta -1)}{q}\, C_2 \dot{a} 
A_2^{\frac{2\delta -1-q}{2}}} \, \right] \;.\label{28}
\end{eqnarray}
The root $R_1$ corresponding to the positive sign in  
eq.~(\ref{28})  
yields a  cosmological horizon. For the physical range of  
parameters $ 0< \delta <<1$ and $ C_2>0$, in an expanding 
universe 
($\dot{a} >0$) the argument of the square root is larger than 
unity and the root $R_2$ corresponding to the negative sign in 
eq.~(\ref{28}) is negative and unphysical. Therefore, we conclude 
that there is no black hole apparent horizon and the  singularity 
at $r=C_2$ (or $R=0$) is naked. 

For completeness, we can consider also a {\em contracting}  
universe 
with 
$\dot{a}<0$, which is obtained for $\delta <0$, although this 
situation is clearly not   interesting for $f(R)$ theories aiming 
at explaining the current  acceleration of the universe and 
$\delta \geq 0$ is required to stabilize the theory against 
explosive local instabilities \cite{mattmodgrav}. In this case  
the negative sign has to be chosen in eq.~(\ref{24}), leading to
\begin{eqnarray}
&& HR^2 + A_2^{\frac{2q^2-2q-4\delta+2\delta q +2}{2q}} \, R 
+\frac{ 
2(q+2\delta-1)}{q} \, C_2 a A_2^{\frac{ 2q^2-2q-4\delta+2\delta 
q+2}{2q}+\frac{2\delta -1-q}{2} } \nonumber\\
&&\nonumber\\
&& =0 \;. \label{29}
\end{eqnarray}
The formal solutions are
\be\label{30}
R_{3,4}=
\frac{A_2^{\frac{2q^2-2q-4\delta+2\delta q+2}{2q}} }{2\left| H 
\right|}
\left[ 1 \pm \sqrt{ 1 - 8 \left| \frac{q+2\delta -1}{q} \right|\, 
C_2 \left| \dot{a} \right| 
A_2^{\frac{2\delta -1-q}{2}}}  \, \right] 
\ee
and are both non-negative.  When the argument of the square root 
is positive the upper sign
yields again a cosmological horizon while the lower sign yields 
a black hole apparent horizon. In the 
solution~(\ref{1})-(\ref{5}) the scale factor $a(t) \approx 
1/t^{|\delta |}$  has a pole-like singularity at  $ t=0$ and 
$\dot{a}$ is always negative, ensuring the existence of the black 
hole apparent horizon at all times. However, as already remarked, 
this situation is completely unphysical.

\section{Quasi-local mass}

The mass of the naked singularity is also of some interest. Due 
to the 
fact that the solution~(\ref{1})-(\ref{5}) 
is not asymptotically flat, the ADM mass is not defined and one 
needs to resort to the concept of quasi-local energy on a 
2-surface 
surrounding the singularity. Thanks to the spherical symmetry of 
this solution it is straightforward to compute the 
Hawking-Hayward \cite{HawkingHayward, Hayward} and the 
Misner-Sharp \cite{MisnerSharp} energies.

By introducing the affine parameters $\xi $ and $\eta$ 
according to 
\begin{eqnarray}
d\xi &=& \frac{1}{\sqrt{2}} \left[ \sqrt{A_2D}\, F 
dt-\frac{1}{A_2^{q/2} C} \sqrt{ 1+\frac{A_2^p H^2 R^2}{C^2 D} } 
\, dR \right] \;, \label{31}\\
&&\nonumber\\
d\eta &=& \frac{1}{\sqrt{2}} \left[ \sqrt{A_2D}\, F 
dt+\frac{1}{A_2^{q/2} C} \sqrt{ 1+\frac{A_2^p H^2 R^2}{C^2 D} } 
\, dR \right] \;, \label{32}
\end{eqnarray}
with 
\be
p=\frac{2q(1-q)+2\delta (2-q)-2}{q} \;,
\ee
the line element~(\ref{18}) is rewritten in the standard form 
\be\label{33}
ds^2=-2d\xi d\eta +R^2 d\Omega^2
\ee
and the Hawking-Hayward quasi-local energy $M_{HH}$ on a 
2-surface  of constant  radius $R$ is given by 
\cite{Hayward}
\be
M_{HH}=R\left( R_{\xi}R_{\eta} +\frac{1}{2} 
\right)=\frac{R}{2}\left( 1-\frac{ A_2^q C^4 D}{C^2 D + A_2^p H^2 
R^2} \right) \;.
\ee

In the presence of spherical symmetry the Misner-Sharp 
quasi-local energy $M_{MS}$ on a  2-sphere $R=$constant is 
defined by \cite{MisnerSharp}
\be
1-\frac{2M_{MS}}{R} =-\nabla^c R\nabla_c R \;,
\ee
which yields
\be
M_{MS}=\frac{R}{2}\left( 1+ \frac{ A_2^q C^4 D}{C^2 D + A_2^p H^2 
R^2} \right) \;.
\ee
In the case of a contracting universe, the two mass notions 
coincide on the black hole apparent horizon $R_{AH}$ given by 
$D=0$, {\em i.e.},  $ M_{HH}=M_{MS}=R_{AH}/2$.

\section{Conclusions}

In view of the fact 
that cosmology  may be showing us the first-ever detected 
deviations from Einstein's gravity  and of the attention given to 
$f(R)$  gravity theories as 
possible models of the cosmic acceleration, it is of great 
interest  to understand black holes and other spherically 
symmetric solutions of $f(R)$ gravity. Since the Jebsen-Birkhoff 
theorem does not hold in these theories, spherically symmetric 
solutions do not have to be static. These theories are designed 
to produce an effective dynamical cosmological constant to 
reproduce the current acceleration of the universe and, 
therefore, dynamical exact solutions 
with spherical symmetry describing a central object embedded 
in a  cosmological background are particularly valuable. 
Unfortunately, this kind of solutions is poorly understood 
already in the context of GR and deserves more  
attention in the future. Few  
examples are available \cite{SultanaDyer, McClureDyer, 
FaraoniJacques,  SaidaHaradaMaeda, Gaoetal, Hideki, 
HidekiHaradaCarr, fate} and further 
work is 
in progress on generalized McVittie solutions 
\cite{VFinpreparation}.  In particular, it seems difficult to 
find {\em 
generic} black holes solutions embedded in cosmological 
backgrounds. The goal of finding interior solutions for 
spherically symmetric $f(R)$ gravity (mainly with numerical 
methods) seems also a worthy one 
\cite{Frolovetc}. All these issues will be addressed in future 
publications.

\section*{Acknowledgements} 
Thanks go to Rituparno Goswami for a discussion and to the 
Natural  Sciences and Engineering Research Council of 
Canada (NSERC) for financial support.

\end{document}